\documentclass[pra, amsmath, amssymb, aps, superscriptaddress, floatfix, twocolumn]{revtex4-2}
\usepackage[T1]{fontenc}
\usepackage[english]{babel}
\usepackage{microtype}
\usepackage[footnotes,definitionLists,hashEnumerators,smartEllipses,hybrid]{markdown}
\usepackage{hyperref}
\usepackage{bm}
\usepackage[dvipsnames]{xcolor}
\usepackage{graphicx}
\usepackage[capitalise]{cleveref}
\usepackage{textcomp}
\usepackage{tabularx, cellspace}
\usepackage{lipsum}
\usepackage{pdfcomment}
\usepackage{bold-extra}
\usepackage{blindtext}
\usepackage{bold-extra}
\usepackage{tabularx}
\usepackage{gensymb}
\usepackage[normalem]{ulem}
\usepackage{pagecolor}

\definecolor{Blue}{RGB}{0, 0, 255}

\hypersetup{
 pdfproducer={LaTeX}, % producer of the document
 pdfkeywords={Version 1.0},
 colorlinks=true,       % false: boxed links; true: colored links
 linkcolor=blue,          % color of internal links (change box color with
 citecolor=blue,        % color of links to bibliography
 filecolor=blue,      % color of file links
 urlcolor=blue
}

\usepackage{titlesec}
\titleformat*{\section}{\bfseries\raggedright\large}
\titlespacing\section{0pt}{10pt plus 2pt minus 2pt}{6pt plus 1pt minus 1pt}
\titleformat*{\subsection}{\bfseries}

\usepackage{caption}
\newcommand{\micron}{$\mu$m }

\newcommand{\jqi}{Joint Quantum Institute, NIST/University of Maryland, College Park, USA}
\newcommand{\pml}{Microsystems and Nanotechnology Division, National Institute of Standards and Technology, Gaithersburg, USA}

\newcommand{\kist}{Center for Opto-Electronic Materials and Devices Research, Korea Institute of Science and Technology, South Korea} %Seoul 136-791,
\newcommand{\cambridge}{Department of Engineering, University of Cambridge, Cambridge, CB3 0FA, United Kingdom}

\begin{document}
\author{Edgar F. Perez}
\email{perezed@umd.edu}
\affiliation{\jqi}
\affiliation{\pml}

\author{Cori Haws}
\affiliation{\cambridge}

\author{Marcelo Davanco}
\affiliation{\pml} 

\author{Jindong Song}
\affiliation{\kist}

\author{Luca Sapienza}
\affiliation{\cambridge}

\author{Kartik Srinivasan}
\email{kartik.srinivasan@nist.gov}
\affiliation{\jqi}
\affiliation{\pml}
\date{\today}

\title{Direct-Laser-Written Polymer Nanowire Waveguides for Broadband Single Photon Collection from Epitaxial Quantum Dots into a Gaussian-like Mode}

\begin{abstract}
    \vspace{-0.2in}
    \noindent     
    Single epitaxial quantum dots (QDs) embedded in nanophotonic geometries are a leading technology for quantum light generation. However, efficiently coupling their emission into a single mode fiber or Gaussian beam often remains challenging. Here, we use direct laser writing (DLW) to address this challenge by fabricating 1~\micron diameter polymer nanowires (PNWs) in-contact-with and perpendicular-to a QD-containing GaAs layer. QD emission is coupled to the PNW's HE$_{11}$ waveguide mode, enhancing collection efficiency into a single-mode fiber. PNW fabrication does not alter the QD device layer, making PNWs well-suited for augmenting preexisting in-plane geometries. We study standalone PNWs and PNWs in conjunction with metallic nanoring devices that have been previously established for increasing extraction of QD emission. We report methods that mitigate standing wave reflections and heat, caused by GaAs’s absorption/reflection of the lithography beam, which otherwise prevent PNW fabrication. We observe a factor of ($3.0~\pm~0.7)\times$ improvement in a nanoring system with a PNW compared to the same system without a PNW, in line with numerical results, highlighting the PNW’s ability to waveguide QD emission and increase collection efficiency simultaneously. These results demonstrate new DLW functionality in service of quantum emitter photonics that maintains compatibility with existing top-down fabrication approaches.

    \vspace*{-0.2in}
\end{abstract}

\maketitle
%\noindent\the\linewidth\\
%\noindent\the\textwidth\\
%\noindent\the\columnwidth

\begin{figure*}[t]
	%\vspace{-.25in}
	\includegraphics[width=1\textwidth]{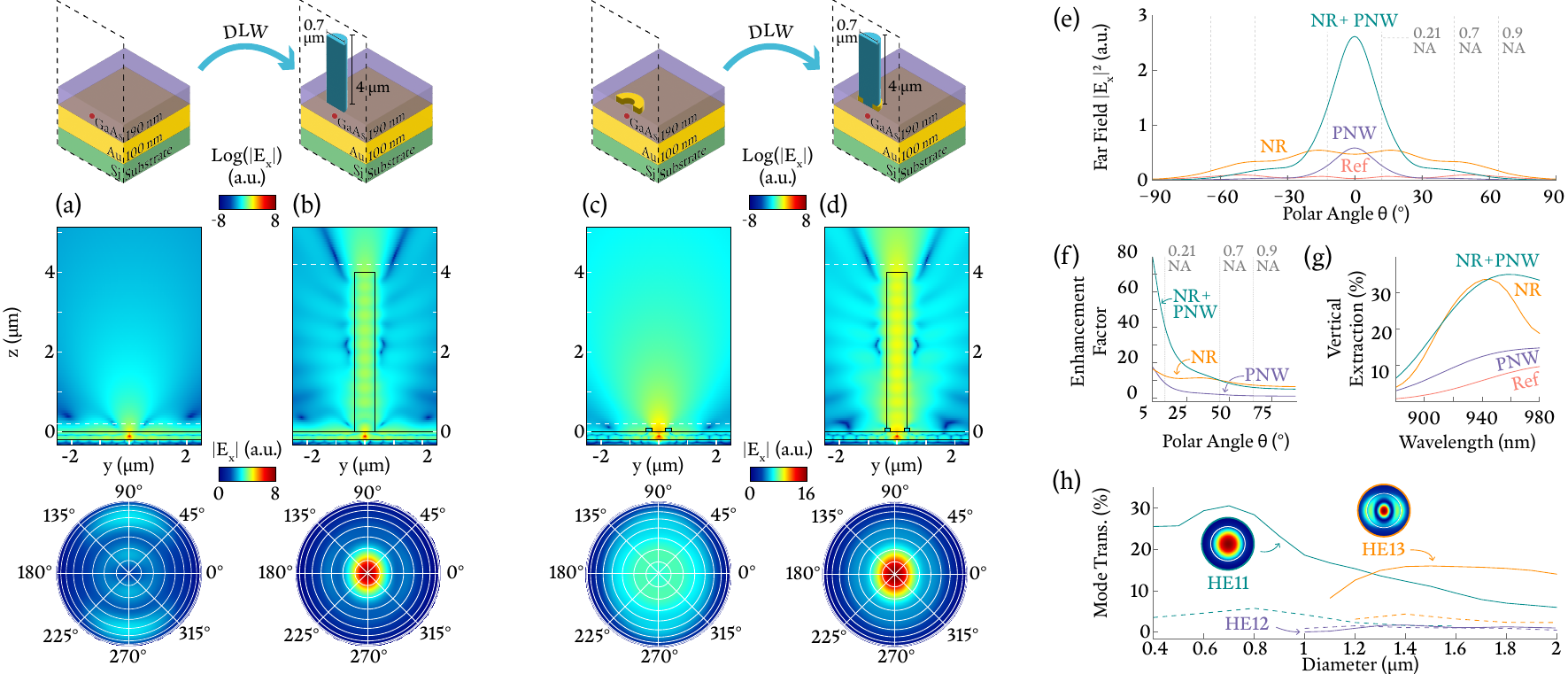}
	%\vspace{.1in}
	\caption{\label{fig:1}
		\textbf{(a-d)} 3D FDTD simulations of systems with and without polymer nanowires (PNWs). For each system (\textbf{a-d}), the top panel is a schematic of the physical device, the middle panel is a cross section of the system's 3D FDTD simulation, and the bottom panel is the far-field projection of the system's emission. The black dashed rectangles in the top panel frame the cross sectional planes displayed in the middle panels, and the white lines in the middle panels indicate the height at which the emission was sampled for far-field projections shown in the bottom panels.
		\textbf{(a)} A reference system, consisting of an x-polarized dipole in the middle of a 190~nm thick GaAs layer on top of an Au back reflector, taken as a starting point for systems in (\textbf{b-d}). 
		\textbf{(b)} A standalone PNW system, with a 0.7~\micron diameter PNW of 4~\micron length, centered over the dipole, showing emission confined- and guided-by the PNW waveguide. 
		\textbf{(c)} A metallic nanoring (NR) system, consisting of an Au NR of inner diameter 236~nm and outer diameter of 437~nm centered above the QD. 
		\textbf{(d)} A metallic NR system with the same dimensions as in \textbf{(c)} and a concentric 0.7~\micron diameter PNW of 4~\micron length (i.e., the NR+PNW system). 
		\textbf{(e)} Line scans ($\phi=90\degree$, y-axis) of the $|E_x|^2$ far-field projections for each system. The PNW (purple) and NR+PNW (green) systems have monotonic and Gaussian-like distributions, whereas the reference (pink) and NR(orange) systems without PNWs have wide and nonmonotonic distributions.
		\textbf{(f)} The ratio of $|E_x|^2$ intensity emitted by the the PNW (pruple), the NR (orange), and the NR+PNW (green) systems compared to the reference system, as a function of polar angle $\theta$. See text for details.
		\textbf{(g)} The transmission of each system through a plane 200~nm above the substrate, showing that all four schemes are broadband. 
		\textbf{(h)} Transmission into the various optical modes of the PNW system (dashed) and the PNW+NR system (solid) as a function of PNW diameter. Insets display the $|E_{x}|$ component of the $HE_{11}$ mode (green) and $HE_{13}$ mode (orange) of a PNW with 1.3~\micron diameter (circumference shown as white line). The $HE_{12}$ mode (purple) of the system is only weakly excited due to poor overlap with an x-polarized dipole. 
	}
\end{figure*}

As fundamental building blocks, single photons (SPs) and entangled photons play key roles in the development of quantum science~\cite{sangouard2012J.Mod.Opt., northup2014NaturePhoton}, and their efficient generation and collection into optical fibers is crucial for quantum technologies such as quantum simulation and quantum communication~\cite{aharonovich2016Nat.Photonics, flamini2018Rep.Prog.Phys.}. Among the many options for SP generation, semiconductor quantum dots (QDs) are a favorable choice, as demonstrated performance has begun to approach the challenging requirements of various applications~\cite{tomm2021Nat.Nanotechnol., wang2019NatPhotonics,senellart2017NatureNanotech, zhai2022Nat.Nanotechnol.}, with top-down nanofabrication techniques, for creating photonic cavities and waveguides that house the QDs, contributing significantly to the success of the platform. However, in many such photonic geometries, realizing high fiber-coupled collection efficiency of SP emission remains an important challenge that, if unaddressed, limits the SP flux delivered to many downstream applications. A viable solution to this problem should ideally increase the collection efficiency without compromising other important performance characteristics, such as SP purity and indistinguishability. To that end, several different efforts to realize such solutions are underway~\cite{bremer2022OE, tomm2021NatNanotech, mantynen2019Nanophotonics}

In recent years, Direct Laser Writing (DLW) has been used to create a variety of three-dimensional (3D) structures that improve fiber-to-chip coupling~\cite{dietrich2018NaturePhoton, lindenmann2012Opt.Express, sartison2021} with increasing levels of automation~\cite{perez2020OE, adao2022OE}, and complexity~\cite{moughames2020Optica}. In the context of SP emission, direct laser written solutions for epitaxially-grown, semiconductor QDs have focused on the fabrication of micro-optics ~\cite{fischbach2017ACSPhotonics, bremer2020APLPhotonics, sartison2021, bremer2022MQT} that shape and collect their far-field radiation. In addition, a variety of DLW structures have been used in collection efficiency enhancement from other single quantum emitters, such as color centers in diamond nanocrystals~\cite{schell_three-dimensional_2013} and organic molecules~\cite{colautti_3d_2020}. In this work, we use DLW to fabricate high-index-contrast cylindrical waveguides, hereafter referred to as polymer nanowires (PNWs), perpendicular to the surface of a semiconductor-based SP emitter substrate. The PNWs operate on the near-field radiation of a SP emitter to directly waveguide-couple its emission to an $HE_{11}$ optical mode that can be well-coupled to a downstream optical fiber. Additionally, the fabrication of PNWs does not require etching of the underlying substrate, which limits potential negative effects on underlying emitters~\cite{liu2018Phys.Rev.Appl.}. As a result, PNWs may operate as standalone devices or in conjunction with existing on-chip solutions - like other broadband collection-enhancing devices or resonant cavities. 

In this work, PNWs are studied as standalone devices and in conjunction with metallic nanorings (NRs), which are broadband collection-enhancing devices \cite{haws2022APL, trojak2017Appl.Phys.Lett.}, to improve the collection efficiency from epitaxially grown InAs/GaAs QDs. PNWs prove to be effective at waveguide-coupling the emission from QDs and we observe a factor of ($3.0~\pm~0.7)\times$ improvement in collection efficiency in a NR system with a PNW compared to the same system without a PNW. The inherently broadband nature of waveguide collection, together with the low dispersion of the materials comprising the PNW, results in improvements in collection efficiency over a large bandwidth. Moreover, the collection into an HE$_{11}$ optical mode is advantageous for direct fiber-coupling applications.

Section I contains a numerical investigation comparing systems with and without PNWs, including the effect of PNW geometry, and the excitation of the various modes of the PNW waveguide. Two key challenges to the fabrication of these devices are standing wave reflections and heating; we present DLW strategies that mitigate these problems in Section II. Optical characterization of the fabricated PNWs is presented in Section III, and we conclude by commenting on the utility that PNWs bring to integrated quantum light sources. PNWs extend the role DLW plays in SP collection by working with near-field emission to directly waveguide-couple SP radiation and increase collection efficiency.

\section{Polymer Nanowire Design}
Polymer nanowires are shown as stand alone devices in Fig.~\ref{fig:1}(a-b), and as complementary devices in Fig.~\ref{fig:1}(c-d). For each system, Fig.~\ref{fig:1}(a-d) provides a schematic of the device (top), a cross section of 3D finite-difference time-domain (FDTD) simulations of the system (middle), and far-field projections of each system's emission (bottom). The dashed black rectangle in the top panels indicates the cross sectional plane displayed in the middle panels. Likewise, the dashed white line in each middle panel indicates the distance from the GaAs substrate at which the emission was sampled for the far-field projections. The device illustrated in Fig.~\ref{fig:1}(a) consists of an x-polarized dipole source (emission wavelengths from 880~nm to 980~nm) embedded in the middle of a 190~nm layer of GaAs with an Au back reflector. This system has been used in recent studies for broadband collection enhancement~\cite{huang2021Adv.Opt.Mater.a, haws2022APL}, and this work takes it as a reference point for understanding the improvements offered by PNWs.

The fundamental concepts of PNW-guided emission can be understood by comparing the reference system in Fig.~\ref{fig:1}(a) to the standalone PNW system in Fig.~\ref{fig:1}(b). The reference system shows in-plane dipole emission that is significantly trapped in the GaAs layer, with the limited free-space emission quickly diverging.
The $|E_{x}|$ component of the far-field projection of the system's emission shows a local minimum at its center and strong radiation at polar angles greater than 40$\degree$, which is not well suited for coupling to the $LP_{01}$ (fundamental, linearly polarized) mode of a single-mode fiber. In contrast, simulating a 4~$\mu$m tall PNW with a 700~nm diameter concentric to the dipole, as shown in Fig.~\ref{fig:1}(b), illustrates that the PNW laterally confines radiation as it propagates through the PNW, along the z-axis, with a far-field distribution that resembles the $LP_{01}$ mode of a fiber. Line scans along the $\phi=90~\degree$ axis (i.e., the y-axis) of the $|E_{x}|^2$ far-field distributions are directly compared in Fig.~\ref{fig:1} (e), which verifies that the distribution of the standalone PNW system (purple) is more Gaussian-like, with a single maximum of greater amplitude at its center compared to the reference system (pink). 

The use of a PNW as a complementary device is studied in Figs.~\ref{fig:1}(c-d), where it is simulated above a metallic NR. As previously reported~\cite{haws2022APL, trojak2018Appl.Phys.Lett., trojak2017Appl.Phys.Lett.}, the NR system shown in Fig.~\ref{fig:1}(c) shows an increase in vertical radiation compared to the reference system, yet displays a local minimum at its center and a large divergence angle. When the NR system is complemented with a PNW (referred to as the NR+PNW system) the emission from the NR is coupled to the optical modes of the PNW. As shown in Figs.~\ref{fig:1}(d-e), this results in a far-field projection with a single maximum at its center that is of greater amplitude than that of the NR system alone. As a complementary device, the PNW has shifted the far-field $|E_x|^2$ distribution towards smaller angles by coupling it to the $HE_{11}$ mode of the PNW. 

% FABRICATION FIGURE
\begin{figure*}[!tb]
	\centering
	\includegraphics[width=.99\textwidth]{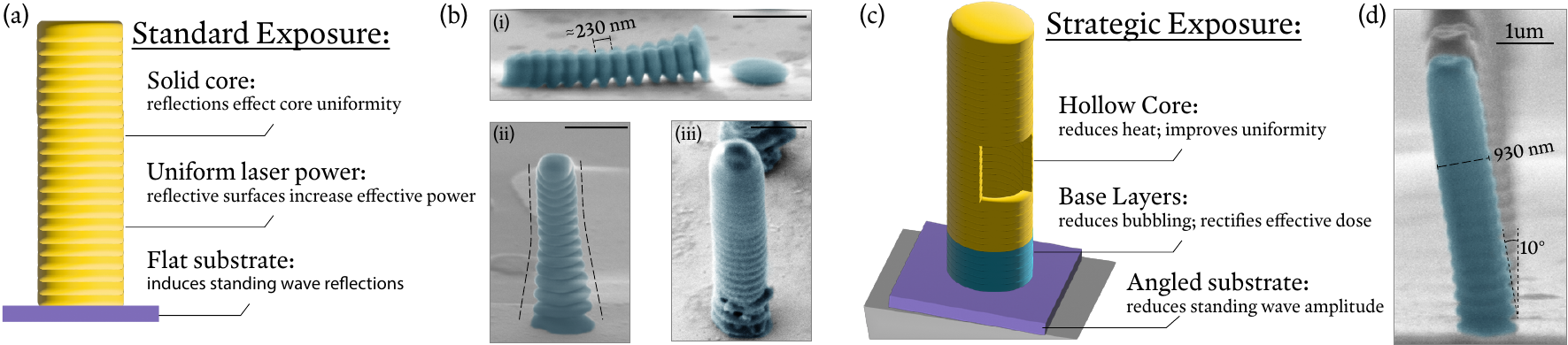}
	\caption{\label{fig:2}
		\textbf{(a)}~A standard approach to DLW structures may use a flat substrate, uniform laser power, and a solid-core design.
		\textbf{(b,i)}~False-color SEM images showing a modulation in the PNW diameter with periodicity of $\approx$230~nm due to standing wave reflections of the pulsed 780~nm lithography beam.
		\textbf{(b, ii)}~Reflections from the substrate also cause a local increase in the effective laser power near the substrate, which increases the diameter of the PNW near the surface.
		\textbf{(b, iii)}~Absorption of the lithography beam further contributes to local heating of resist, which can lead to deformations and resist vaporization (e.g., "bubbling").
		\textbf{(c)}~The exposure strategy used in this work with a substrate mounted at a 10$\degree$ angle, reduced-power base layers, and a hollow core. 
		\textbf{(d)}~False-color SEM images showing PNWs with improved uniformity and decreased corrugation amplitudes realized with the proposed exposure strategy. The 3D flexibility of DLW is used to correct for the angle induced by the angled substrate in \textbf{(d)} (see Fig.~\ref{fig:3}), but is intentionally included in this device for illustrative purposes. All scale bars correspond to 1~\micron. 
	}
\end{figure*}

For each of the PNW, NR, and NR+PNW systems, Figure~\ref{fig:1}(f) plots the ratio of the integrated electric field intensity within a solid polar angle $\theta$, calculated as $\int{|E_x|^2 d\Omega}$, in the system to that of the reference system. As expected, the PNW system (purple) and the NR+PNW system (green) increase the radiation at small angles. At large angles, the total extracted power of the PNW systems approaches the total extracted power of the underlying system, suggesting that the PNWs have limited effect on the extraction efficiency of the underlying devices. In other words, the PNWs operate on the near field emission of the QD systems to increase collection efficiency without significantly modification to their performance. For $\theta \approx 12 \degree$, the divergence angle of SM980 fiber~\cite{NIST_disclaimer}, the PNW (NR+PNW) system shows an increase of approximately 8.5$\times$ (41.0$\times$) in collection efficiency compared to the reference system. For the same angle, the NR+PNW system shows approximately 3.3$\times$ the collection efficiency of the NR system without a PNW (orange curve). Finally, Fig.~\ref{fig:1}(g) compares the broadband performance of each system, and shows that PNWs (purple) are broadband devices, making them compatible with NR systems (orange), endowing the NR+PNW system (green) with broadband performance as well. 

Like most waveguides, the optical properties of a PNW will depend on its geometry. Figure~\ref{fig:1}(h) shows the simulated transmission of dipole radiation into various optical modes of the PNW as a function of nanowire diameter $D$ for the standalone system (dashed) and the NR+PNW system (solid). In general, the two systems display similar behavior even with the collection from the NR+PNW system being approximately 4$\times$ to 6$\times$ greater. As expected, there is an optimal diameter $D\approx700$~nm at which coupling into the fundamental $HE_{11}$ mode of the PNW is maximized, with a monotonic decrease for $D>700$~nm. This monotonic decrease is {not associated with the excitation of higher order modes, so it is} attributed to a mismatch in the PNW mode profile and the emission of the underlying device, meaning that it may be possible to design a PNW in conjunction with underlying devices to prevent this decrease. When $D\approx1.3$~$\mu$m, coupling to the $HE_{13}$ mode of the PNW begins to dominate and coupling to the $HE_{11}$ mode is reduced by half. We note that coupling to the $HE_{12}$ mode of the PNW is inefficient at all diameters due to poor overlap with an x-polarized dipole.

In summary, numerical studies of PNWs support their use as interfaces for direct waveguide-coupling of dipole emission from solid-state SP emitters. In both standalone and cooperative systems the PNWs offer waveguiding, $HE_{11}$ emission profiles, and significant farfield enhancement for small emission angles $\theta \lessapprox20\degree$. QD emission couples to the fundamental mode of PNWs for $D \le 1$~\micron, with an ideal diameter of $700$~nm. However, beyond $700$~nm, the large degree of overlap between the $HE_{11}$ mode of the PNW and the $LP_{01}$ mode of a single-mode fiber, together with the small divergence of PNW emission, may still provide an overall increase in the fiber collection efficiency of a system.

\section{Fabrication of PNWs}
The direct laser writing used in this study employs strongly focused 780~nm femtosecond laser pulses to polymerize a small voxel of photoresist through two photon excitation. Despite the minimum resolution of DLW systems routinely reaching sub-micron scales~\cite{zandrini2019OME, sun2003J.Light.Technol., purtov2019Nanomaterialsa}, the DLW of sub-micron features requires special attention to proximity effects during exposure~\cite{saha2017JournalofMicroandNano-Manufacturinga} and structural deformations during development~\cite{purtov2018MicroelectronicEngineeringc}. Near the surface of reflective and absorptive substrates like GaAs, additional challenges arise. The partial reflection of the high intensity lithography pulses induces standing waves that lead to modulations of the effective exposure dose. Furthermore, the partial absorption of the pulses causes local heating of the substrate, resist, and any underlying structures, which affects the polymerization threshold of the resist~\cite{harnisch2019Opt.Mater.Express}. Each of these effects deteriorate the quality of small features near a substrate, and realizing sub-micron diameter PNWs requires their mitigation.

A standard approach to DLW may use a flat substrate, uniform laser power, and a solid-core design in the fabrication of PNWs, like that illustrated in Fig.~\ref{fig:2}(a). Fabrication outcomes from the use of this scheme with IP-Dip photoresist~\cite{NIST_disclaimer} are shown in Fig.~\ref{fig:2}(b, i-iii), where unmitigated reflections cause standing-wave-induced deep corrugations that can separate PNWs at their base (i), an increase in the diameter of the PNW near its base (ii) (due to increased effective dose), and a shift of the polymerization/damage thresholds that can cause vaporization of the resist (e.g., "bubbling") near the substrate (iii). As shown in Fig.~\ref{fig:2}(b,i) the period of the corrugations are approximately~230~nm, which roughly correlate with the expected period of~260~nm for a~780~nm continuous wave laser in IP-Dip. 

To mitigate heat generation and reflections, we employed the exposure strategy shown in Fig.~\ref{fig:2}(c). In this strategy, a custom mount positions the GaAs substrate at a 10$\degree$ angle relative to the lithography beam to limit the overlap between incident and reflected pulses; base layers of lower lithography power rectify variations in effective dose; and a hollow-core design reduces local heating by limiting the total dose required to fabricate the PNW. After exposure, samples were UV-cured in a solvent bath to polymerize their cores and prevent structural deformation of the structures~\cite{purtov2018MicroelectronicEngineeringc}. The scanning electron micrograph (SEM) shown in Fig.~\ref{fig:2}(d) demonstrates the successful fabrication of a PNW at the 10$\degree$ angle provided by the custom mount. In all subsequent fabrication runs, a~$-10\degree$ correction was applied to the PNW design to compensate for the $+10\degree$ angle introduced by the mount, which is possible due to the 3D flexibility of DLW. The angled PNW shows greatly reduced corrugation amplitudes and improved uniformity compared to those in Fig.~\ref{fig:2}(b). 

\begin{figure}[!tb]
	\includegraphics[width=\columnwidth]{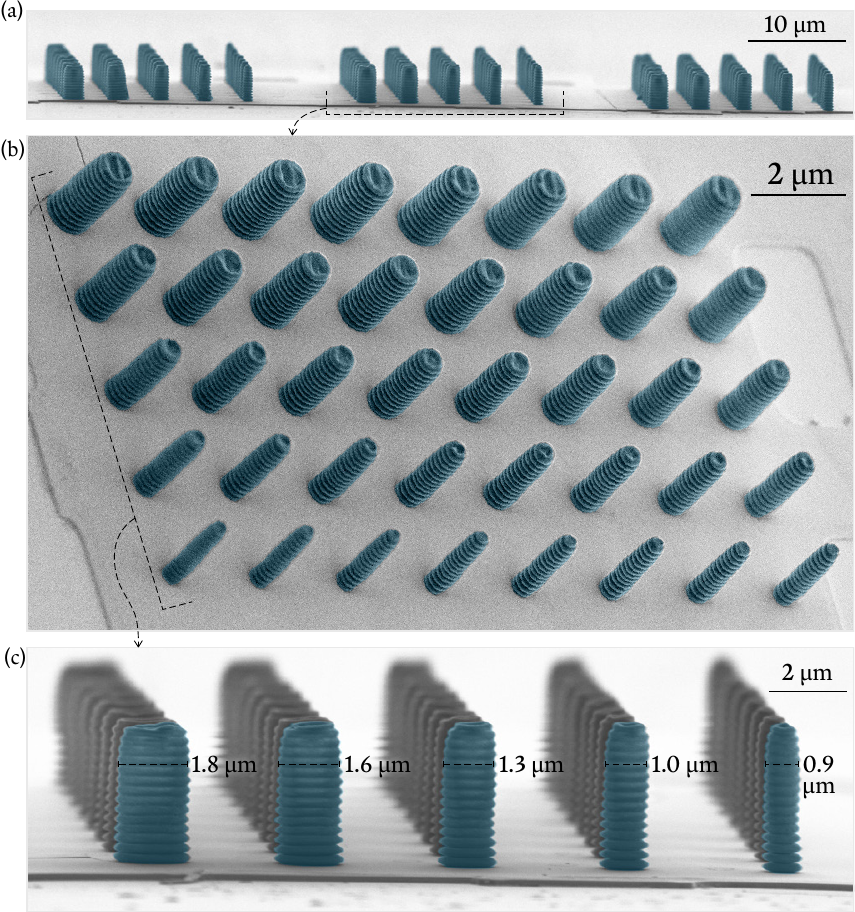}
	\caption{\textbf{(a)}~False-color SEM images of PNWs fabricated on GaAs substrates with Au back reflectors with 100~\% yield using the strategy proposed in Fig.~\ref{fig:2}(c). \textbf{(b)}~A perspective view of the central set of devices in (a), showing straight-standing PNWs of different diameters. \textbf{(c)}~A lateral view of a PNW array, showing greatly reduced sidewall corrugations at all diameters compared to those fabricated using a standard strategy (Fig.~\ref{fig:2}(b)).}
	\label{fig:3}
\end{figure}

To validate the general exposure strategy proposed in Fig.~\ref{fig:2}(c), a total of 120 PNWs of varying diameters were fabricated on the reference substrate shown in Fig.~\ref{fig:1}(a). The nanowires were fabricated using IP-Dip photoresist~\cite{NIST_disclaimer}, 6.25~mW to 12~mW of time-averaged laser power, with 80~fs to 100~fs pulses focused by a numerical aperture (NA) 1.4 objective at an 80~MHz repetition rate and a laser scan speed of 1~mm/s. As shown in Fig.~\ref{fig:3}(a-b), the fabrication had a 100 percent yield of angle-corrected (i.e., straight-standing and perpendicular to the substrate) PNWs with reduced corrugation depths. Figure~\ref{fig:3}(c) shows PNWs with diameters varying from $0.9$~\micron to $1.8$~\micron, showing the uniformity of each design with respect to diameter. Despite the Au back reflector in the reference design providing very strong reflections, the angled substrate reduced the depth of the corrugations to $\approx 150$~nm for the reference system and $\approx 50$~nm for PNWs printed on GaAs with no back reflector (Fig.~\ref{fig:2}(d)).

\begin{figure}[tb]
	\includegraphics[width=\columnwidth]{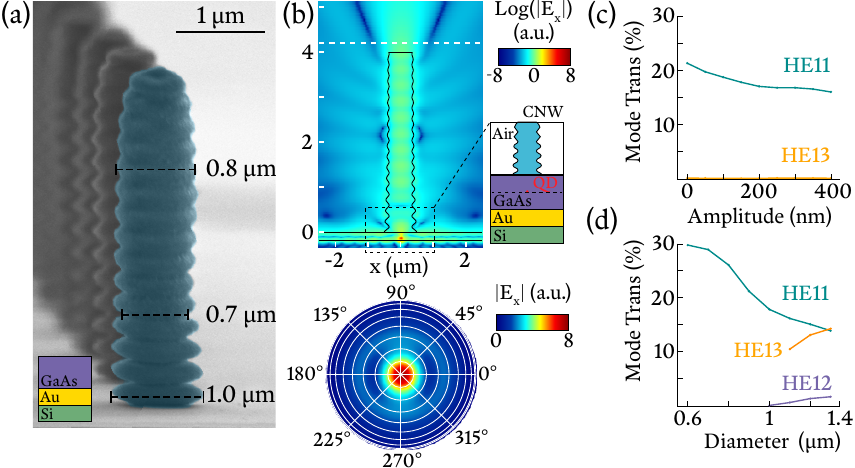}
	\caption{\textbf{(a)} Close up scanning electron micrograph of a PNW on a GaAs substrate with an Au back reflector and showing an $\approx$~150~nm corrugation amplitude. \textbf{(b)} FDTD simulation of a corrugated NW showing that light is still confined in a PNW mode propagating along the z axis (top) and far-field emission is still azimuthally symmetric and Gaussian-like. \textbf{(c)} The coupling efficiency into the $HE_{11}$ modes of a 1~\micron diameter PNW  as a function of corrugation amplitude. \textbf{(d)} The excitation of higher order modes for a fixed corrugation amplitude but varying PNW diameter, showing similar performance to the uncorrugated device in Fig.~\ref{fig:1}(h).}
	\label{fig:4}
\end{figure}

Despite the progress on PNW fabrication, the persistence of the corrugations warrants a numerical understanding of their influence on device performance. Figure~\ref{fig:4}(a) shows the fabrication of a PNW on the reference substrate (inset) with various dimensions highlighted to show diameter variation along the z axis. In particular, the device shows corrugation amplitudes that decrease as a function of the distance $z$ from the substrate surface. Corrugations were introduced into FDTD simulations as a variation of the diameter $D(z) = D_0 + \delta(z)$ with $\delta(z) = \frac{2A}{1+\gamma z} cos^2(z \frac{\pi}{\lambda_c})$, where $A$ is the corrugation amplitude, $\gamma$ linearly decreases the amplitude for increasing $z$, and $\lambda_c$ is the corrugation wavelength. Figure~\ref{fig:4}(b) shows an FDTD simulation for a PNW on the reference substrate with $D_0=0.7$~\micron, $A=150~$nm, $\gamma=0.75$, and $\lambda_c=230$~nm. As shown in the simulation, the corrugated device still confines emission to the z-axis and has a Gaussian-like far-field distribution amenable to fiber-coupling. The effect of the corrugation amplitude is investigated in Fig.~\ref{fig:4}(c), which shows that the corrugation depth monotonically reduces the coupling efficiency from the dipole to the $HE_{11}$ mode of the PNW. Even so, for $A = 200$~nm the corrugations introduce $<20~\%$ loss compared to the ideal PNW with $A=0$~nm. The modulations tend to shift the average diameter of the PNW, so that a small shift in the optimal waveguide dimensions is expected. Fixing the corrugation depth to $150$~nm depth, like that of the fabricated device in Fig.~\ref{fig:4}(a), and varying $D_0$ (Fig.~\ref{fig:4}(d)) verifies that the optimal diameter decreases to $\approx 600$~nm. In general, the corrugations reduce the optimal PNW diameter and the HE$_{11}$ mode transmission, but they do not prevent the PNW from confining radiation and emitting a directional gaussian-like mode.

\section{Optical Characterization of PNWs}
\begin{figure*}[tb]
	\includegraphics[width=\textwidth]{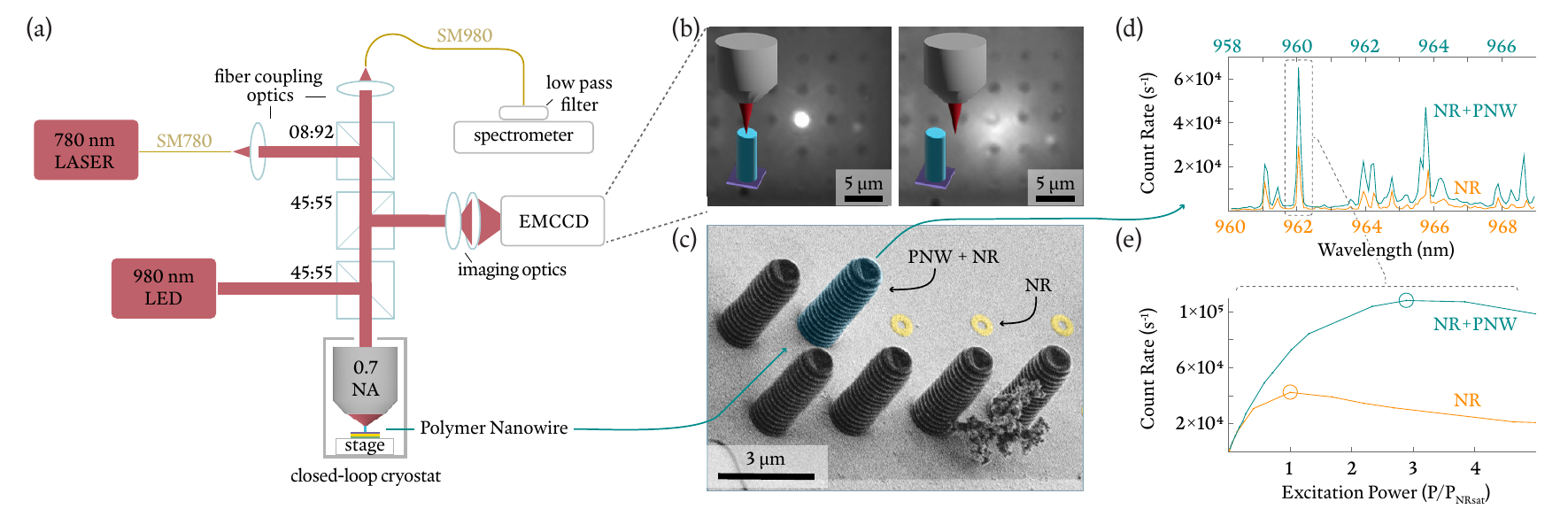}
	\caption{\label{fig:5}
        \textbf{(a)} Schematic of the micro-photoluminescence setup used for PNW characterization (see main text for details). 
        \textbf{(b)} Images captured by the EMCCD camera when an excitation beam is aligned-to and confined by a PNW (left) and when the beam is laterally displaced from that PNW (right) showing beam divergence. Substrate illumination in this image is provided by an incoherent LED. A PNW array makes up the dark circles in the images. 
        \textbf{(c)}~False color SEM image of 9 NRs (yellow) with PNWs fabricated on top of 6 of the NR locations. Data from the highlighted NR+PNW is presented in panels (\textbf{d-e}) before and after PNW fabrication. \textbf{(d)} Emission spectra collected from the test site indicated in panel (c) before (i.e., a NR only system) and after the fabrication of the PNW (i.e., a NR+PNW system), showing a broadband spectral enhancement of emission lines in the NR+PNW system over a span of 9~nm. Spectral shift discussed in main text. \textbf{(e)} Power series for the highest intensity peak of the NR system (orange) and the NR+PNW system (green), showing a factor of $\approx2.6\times$ enhancement at QD saturation. The spectra shown in \textbf{(d)} correspond to the circled data points in \textbf{(e)}.
	}
\end{figure*}

Implementing the fabrication strategy proposed above, PNWs were fabricated on both the reference structure (shown in Fig.~\ref{fig:3}) and on the NR devices, as shown in Fig.~\ref{fig:5}(c). To avoid spurious reflections from the Au NR, the PNWs fabricated on NRs had $D_0 \approx 1.4$~$\mu$m, which reduces the PNW-to-NR coupling efficiency, as shown in Fig.~\ref{fig:4}(d), but provides an emission profile that overlaps well with the $LP_{01}$ mode of a single mode fiber. The standalone PNW and the NR+PNW systems were optically characterized using the micro-photoluminescence set up shown in Fig.~\ref{fig:5}(a). The devices were cooled to a temperature of $\approx$ 4 K in a closed-loop cryostat, and illuminated using a 980~nm light emitting diode (LED) or a 780~nm tunable laser. The laser was operated with continuous-wave emission for above-band QD excitation and the LED provided broadband wide-field illumination of the substrate. The laser source is focused onto the sample by a 0.7 NA objective to a spot approximately 1.4~\micron in diameter. On the other hand, the LED source is collimated on the sample to provide wide-field illumination. During data acquisition, only the excitation beam was active, i.e., the widefield LED illumination was turned off. The focal plane of the objective is at the top of PNWs (NRs) for measurements of PNW (NR) emission, and the collected signal is routed to an electron multiplying charge-coupled device (EMCCD) camera and a 0.5~m long grating spectrometer through a series of beamsplitters with splitting ratios and positions as indicated in Fig.~\ref{fig:5}(a). 

Figure~\ref{fig:5}(b) shows a representative image captured by the EMCCD when a PNW array is illuminated by the LED and the laser simultaneously. The objective is focused on the top plane of a PNW array, and each PNW appears as a small darkened dot in the image. Using a piezo stage within the cryostat, the excitation can be aligned to a PNW, the result of which is shown in the left panel of the figure. In this image, the laser appears as a bright illuminated circle. On the other hand, displacing the excitation beam in a lateral direction produces the image shown in the right panel, where the excitation beam is not confined by the PNW, thus diverging to a large out-of-focus spot spread out on the substrate surface. These images demonstrate that the PNWs are functioning as waveguides perpendicular to the substrate surface. 

In order to quantify the impact of PNWs on a QD's emission, the same QD should be measured before and after the fabrication of a PNW. The metallic NR platform is well suited for this comparison, given that the centers of the rings provide a well-defined location for optical measurements before and after PNW fabrication. Figure~\ref{fig:5}(c) shows a set of six PNWs printed directly on an array of metallic NRs, which were fabricated on a GaAs substrate with a high density of emitters. As a representative device, the PNW highlighted by the green arrow, which is centered over a NR, was studied before and after fabrication of the PNW. Figure~\ref{fig:5}(d) shows the emission spectra of this NR before (orange) and after (green) the fabrication of the PNW. The top and bottom axes of this figure correspond to the emission wavelengths for each system, and we note that the spectra of the NR+PNW system was shifted by $\approx$ 2~nm relative to that of the NR alone. This spectral shift may be due to polymer-induced strain on the substrate from possible shrinkage of the cold DLW polymer~\cite{bremer2020APLPhotonics, sartison2017SciRepa}. Nevertheless, the emission lines can be uniquely identified by the structure of the spectra over a wide bandwidth, as shown in the figure. The spectra in Fig.~\ref{fig:5}(d) were taken near QD saturation, at powers indicated by the open circles in Fig.~\ref{fig:5}(e). We observe a broadband spectral enhancement of QD emission in the NR+PNW system compared to the NR system alone.

As a representative single-QD comparison, the emission of the spectral line at $\approx$~962~nm was studied for increasing amounts of excitation power, as shown in Figure~\ref{fig:5}(e) in units of NR saturation power. At saturation, the maximum count rates measured at the spectrometer were $\approx$~42000~s$^{-1}$ for the NR system and $\approx$~108000~s$^{-1}$ for the NR+PNW. This observed enhancement factor of 2.6$\times$ is inline with the $\approx$3.3$\times$ enhancement predicted in Fig.~\ref{fig:1}(f) for an SM980 (NA = 0.21) fiber-coupled input to the spectrometer. We note however, that the count rate enhancement is achieved at an excitation power 3$\times$ greater than that of the NR device alone. The PNW was designed with a 1.4~\micron diameter to prevent spurious reflections at the ring's surface during fabrication but at the cost of becoming a multi-mode waveguide for light at 950~nm, as shown in Fig.~\ref{fig:1}(h). Thus, the 780~nm excitation beam likely couples to an even greater number of modes, some of which may not excite the QD in the NR configuration with the same efficacy as the free-space beam in the NR configuration. In principle, the excitation efficiency can be optimized by using a PNW that is single mode at both the excitation and emission wavelengths, but in this study the minimum PNW diameter is restricted to avoid the NR reflections. On average, we observe a ($3.0~\pm~0.7)\times$ enhancement over the spectral window shown in Fig.~\ref{fig:5}(e), where the error represents the standard deviation of five prominent spectral lines and the leading source of error is likely the distribution of the underlying QDs near the principal axis of the PNW. In general, the PNW has enhanced the fiber-coupling of the NR platform by directly waveguide-coupling its emission to the $HE_{11}$ optical mode, and we expect further improvements to be possible through the use of modified PNW designs (e.g., elliptical PNWs), modified NRs (e.g., optimized for PNW coupling), different underlying devices (e.g., circular Bragg gratings~\cite{davanco_circular_2011,liu2019Nat.Nanotechnol., sapienza2015NatCommun}), or individually pre-positioned QDs.

\section{Discussion and context}

We have demonstrated that DLW can be used to fabricate PNWs that directly waveguide-couple the emission from InAs/GaAs QDs while increasing collection efficiency into a down-stream fiber. Using numerical simulations, we show that dipole emission can be coupled to the optical modes within a PNW, which is expected to increase the collection efficiency by a down-stream single mode fiber due to the large overlap of a PNW's $HE_{11}$ mode with the $LP_{01}$ mode of a single mode fiber. We have also shown that PNWs are compatible with other on-chip devices like metallic NRs, since they do not require etching or modification of the underlying substrate. PNWs were fabricated on reflective substrates with DLW by suppressing standing wave reflections and mitigating heat generation. Comparing a NR system before and after the fabrication of a PNW, we observe an approximately three-fold increase in the collection of the saturated flux of a single QD emission line when coupling the excitation light through the PNW. Our realization of wavelength-scale PNWs near the surface of a substrate that is both reflective and absorptive brings new functionality to the expanding field of 3D quantum photonic circuitry, and we expect that advances in DLW fabrication, PNW design, and modified on-chip devices will continue to improve the performance of PNW-enhanced systems. 

This investigation has focused on azimuthally symmetric PNWs that are perpendicular to the device substrate, but a broadband PNW (or NR+PNW), like those in this study, also offers full 3D flexibility. This may make PNWs well suited for use with other solid state devices, like micro-pillars and circular Bragg gratings. For example, the confined photons can be rerouted to other on-chip elements as in various examples of photonic wire-bonding~\cite{lindenmann2012Opt.Express}. In particular, PNWs may be especially well suited for this function when used in conjunction with III-V nanowires~\cite{mantynen2019Nanophotonics}, where the $HE_{11}$ mode of the III-V nanowire can be well coupled to the $HE_{11}$ mode of the PNW, which then routes the SP emission. We further note that the use of anti-reflection coatings (at the DLW lithography wavelength) on the surface of the GaAs may further improve PNW quality, as would the use of a distributed Bragg reflection mirror as the system's back reflector (at the cost of reduced reflection bandwidth). Finally, other possibilities include the use of elliptical nanowires for polarization-maintaining emission, or tapered PNWs that adiabatically transfer power from the higher order modes to the fundamental $HE_{11}$ mode for further-improved performance.

%Bibliography
\bibliographystyle{naturemag}
\bibliography{Biblio}

\begin{thebibliography}{10}
\bibitem{NIST_disclaimer}
\bibinfo{note}{Certain commercial products or names are identified to foster
  understanding. Such identification does not constitute recommendation or
  endorsement by the National Institute of Standards and Technology, nor is it
  intended to imply that the products or names identified are necessarily the
  best available for the purpose.}
\end{thebibliography}

\medskip
\noindent \textbf{Data availability} The data that supports the plots within this paper and other findings of this study are available from the corresponding authors upon reasonable request.

\medskip
\noindent \textbf{Acknowledgments} This work is supported by the NIST-on-a-chip program and the Leverhulme Trust, Grant IAF-2019-01. J.D.S. would like to acknowledge the partial support received from the IITP grant, which is funded by the Korean government (MSIT 2020-0-00841).

\medskip
\noindent \textbf{Author contributions} E. P. led the project, PNW design and fabrication, and conducted experiments. C.H. conducted optical characterization experiments, contributed to theoretical understanding and data analysis. M. D. contributed to theoretical understanding, the design of the PNWs, and the interpretation of data. J. D. S. grew the high density InAs/GaAs QDs. L. S. contributed to the theoretical understanding of PNWs, their design, and the fabrication of devices. K. S. contributed to the theoretical understanding of PNWs, standing wave mitigation, and data analysis. E. P. and K. S. prepared the manuscript. All authors contributed and discussed the content of this manuscript.
\clearpage

\end{document}